\documentclass[aps,twocolumn,groupedaddress]{revtex4}

\usepackage[oztex]{graphicx}

\begin{document}
\bibliographystyle{apsrev}

\preprint{LA-UR-00-2319}

\title[Na$^+$(aq)]{The hydration number of Na$^+$ in liquid water}



\author{Susan B. Rempe}
\affiliation{Theoretical Division, Los Alamos National Laboratory, Los
Alamos, New Mexico 87545 USA}

\author{Lawrence R. Pratt}
\affiliation{Theoretical Division, Los Alamos National Laboratory, Los
Alamos, New Mexico 87545 USA}


\date{\today}

\begin{abstract}
An `ab initio' molecular dynamics simulation of a Na$^+$ ion in
aqueous solution is presented and discussed.  The calculation treats a
Na$^+$ ion and 32 water molecules with periodic boundary conditions on
a cubic volume determined by an estimate of zero partial molar volume
for this solute in water at normal density and  at a temperature of
344$\pm$24~K. Analysis of the last half of the 12 ps trajectory shows
4.6 water molecules occupying the inner hydration shell of the Na$^+$
ion on average, with 5 being the most probable occupancy. The
self-diffusion coefficient observed for the Na$^+$ is
1.0$\times$10$^{-5}$~cm$^2$/s. The quasi-chemical theory of solutions
provides the framework for two more calculations. First a
complementary calculation, based on electronic structure results for
ion-water clusters and a dielectric continuum model of outer sphere
hydration contributions, predicts an average hydration shell occupancy
of 4.0.  This underestimate is attributed to the harmonic
approximation for the clusters in conjunction with the approximate
dielectric continuum model treatment of outer sphere contributions.
Finally, a maximum entropy fitting of inner sphere occupancies that
leads to insensitive composite free energy approximations suggests a
value in the neighborhood of -68~kcal/mol for the hydration free
energy of Na$^+$(aq) under these conditions with no contribution supplied for
packing or van der Waals interactions.

{\it keywords:}  ab initio molecular dynamics,
dielectric continuum, electronic structure, hydration, information
theory, quasi-chemical theory, sodium ion
\end{abstract}
\pacs{}

\maketitle


\section{Introduction}
Solvation of simple ions in aqueous solution is not yet fully
understood despite its fundamental importance to chemical and
biological processes. For example, disagreement persists regarding the
hydration number of the Na$^+$ ion in liquid water. A pertinent
problem of  current interest centers on the selectivity of biological
ion channels; it seems clear that the selective transport of K$^+$
relative to Na$^+$ ions in potassium
channels\cite{kchannel:98,guidoni:99,laio:99} depends on details of
the ion hydration that might differ for K$^+$ relative to Na$^+$.

Experimental efforts to define the hydration structure of Na$^+$(aq)
using diffraction\cite{caminiti:80,skipper:89} and
spectroscopic\cite{Michaellian:78} methods produce a hydration number
ranging between four and six\cite{Ohtaki:93}. Simulations have
obtained a range of values, but most predict six water molecules in
the inner hydration sphere of the Na$^+$ ion%
\cite{Heinzinger:79,Mezei:81,Impey:83,Chandrasekhar:84,%
bounds:85,wilson:85,Zhu:91,Heinzinger:93,lee:94,toth:96,obst:96,%
Koneshan:98b}.  An `ab initio' molecular dynamics simulation produced
five inner shell  water molecules neighboring
Na$^+$(aq)\cite{schwegler:00}.

An important limitation of theoretical studies of ion hydration
concerns the sufficiency of model force fields used in classical
statistical mechanical calculations.  In the most customary
approaches, interatomic force fields used in theories or simulations
derive from empirical fits of a parameterized model to a variety of
experimental data. `Ab initio' molecular dynamics avoids this
intermediate modeling step by approximate solution of the electronic
Schroedinger equation for each configuration of the
nuclei\cite{marx:99,alfe:2000}. This technique thus goes significantly
beyond conventional simulations regarding the accuracy of the force
fields. It also augments  theories built more directly on electronic
structure studies of ion-water complexes by adopting approximate
descriptions of the solution environment of those
complexes\cite{rempe:99,pratt:98,feature,martin:97,pratt:99,G98}.

Relative to conventional simulations, `ab initio' molecular dynamics
simulations also have some important limitations due to the high
computational demand. Applications of the method have been restricted
to small systems simulated for short times. For example, an `ab
initio' molecular dynamics study\cite{schwegler:00} of the Na$^+$(aq)
ion comparable to the present work obtained a thermal trajectory
lasting 3 ps after minimal thermal aging. The present work, though
still limited to relatively small systems, pushes such calculations to
longer times that might permit more precise determination for
Na$^+$(aq) of primitive hydration properties. The analysis here
utilizes the last half of a 12~ps thermal trajectory.  The
quasi-chemical
theory\cite{rempe:99,pratt:98,feature,martin:97,pratt:99} and separate
electronic structure calculations on Na(H$_2$O)$_n{}^+$ complexes
assist in this analysis.

\begin{figure}
\begin{center}
\leavevmode
\includegraphics{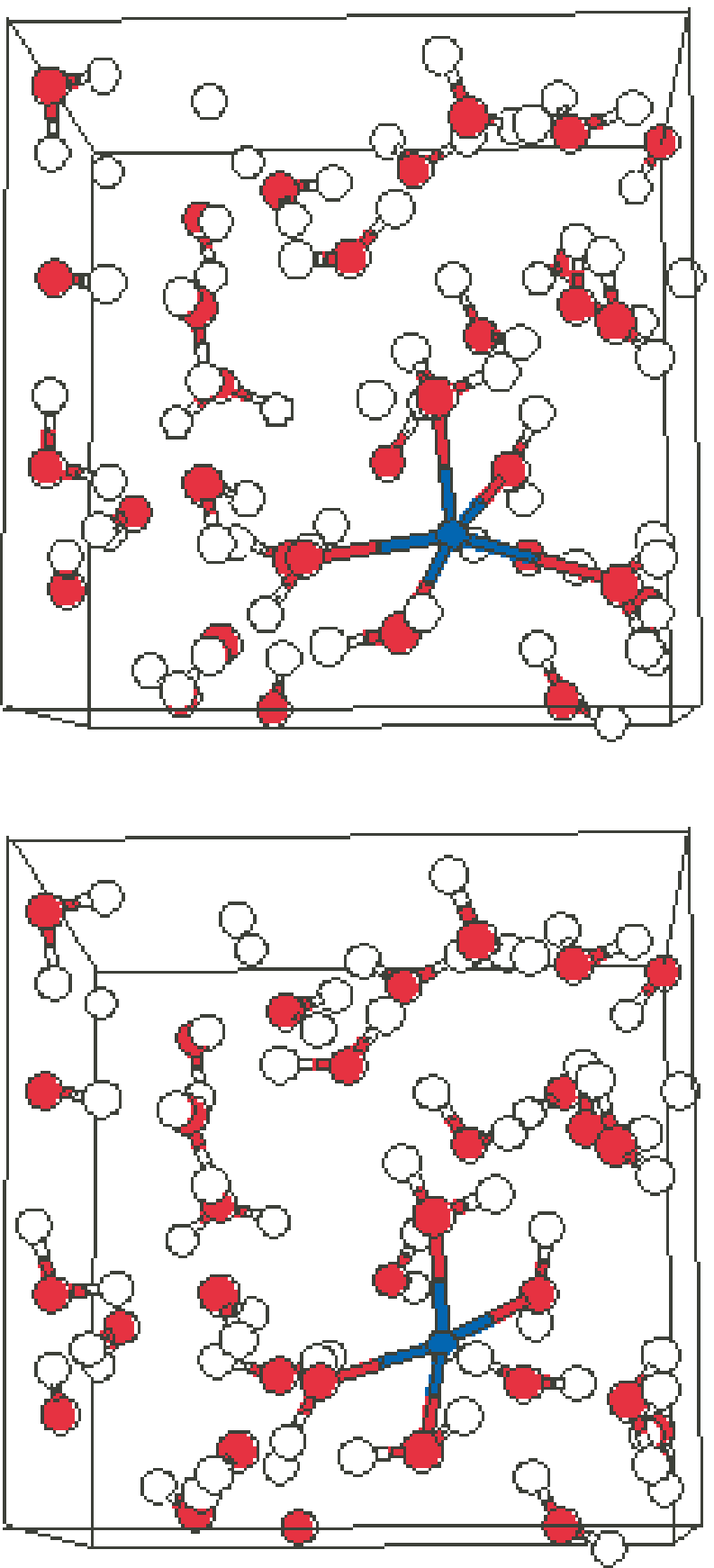}
\end{center}
\caption{Structures from `ab initio' molecular dynamics calculations.
In the top panel, the Na$^+$ ion has five (5) inner shell water
molecule neighbors. The bottom panel shows the four-coordinate
structure produced 70~fs later.  The bonds identify water oxygen atoms
within 3.1~\AA\ of the Na$^+$ ion.  The hydrogen, sodium, and oxygen
atoms are shown as open, black, and gray circles, respectively.}
\label{sim:fig}
\end{figure}

\section{Methods}

The system consisted of one Na$^+$ ion and 32 water molecules in a
cubic box with edge 9.86518~\AA\ and periodic boundary conditions. The
dimensions of the box correspond to a water density of 1 g/cm$^3$ and
zero partial molar volume for the solute. Initial conditions were
obtained as in  an earlier `ab initio' molecular dynamics simulation
on Li$^+$(aq)\cite{rempe:99}. In that earlier work, an optimized
structure  for the inner sphere Li(H$_2$O)$_6{}^+$ complex was
equilibrated with 26 water molecules under conventional simulation
conditions for liquid water, utilizing a current model force field and
assuming a partial molar volume of zero. In the present calculation,
the same pre-equilibrated system was used as an initial configuration
for the `ab initio' molecular dynamics except that an optimized
structure for the inner sphere Na(H$_2$O)$_6{}^+$ complex replaced the
hexa-coordinated Li$^+$ structure. Constant pressure or constant water
activity simulations, defined by intensive rather than extensive
variables, probably would produce a more useful characterization of
the solvent thermodynamic state for these small systems, but those
alternatives are currently impractical.

Molecular dynamics calculations based upon a gradient-corrected
electron density functional description of the electronic structure
and interatomic forces were carried out on this Na$^+$(aq) system
utilizing the VASP program\cite{vasp1}. The ions were represented by
ultrasoft pseudopotentials\cite{vasp2} and a kinetic energy cutoff of
31.5 Rydberg limited the plane wave basis expansions of the valence
electronic wave functions.  The equations of motion were integrated in
time steps of 1~fs, which is small enough to sample the lowest
vibrational frequency of water. A thermostat constrained the system
temperature to 300~K during the first 4.3 ps of simulation time. After
removing the thermostat, the temperature rose slightly and then
leveled off by 6 ps to an average of 344 $\pm$ 24~K. During the
simulation, the initial $n$=6 hydration structure relaxed into $n$=4
and $n$=5 alternatives, such as those shown in Fig.~\ref{sim:fig}. All
analyses reported here rely on the trajectory generated subsequent to
the 6 ps of aging with the system at a temperature elevated from room
temperature.

\begin{figure}
\begin{center}
\leavevmode
\includegraphics[scale=0.6]{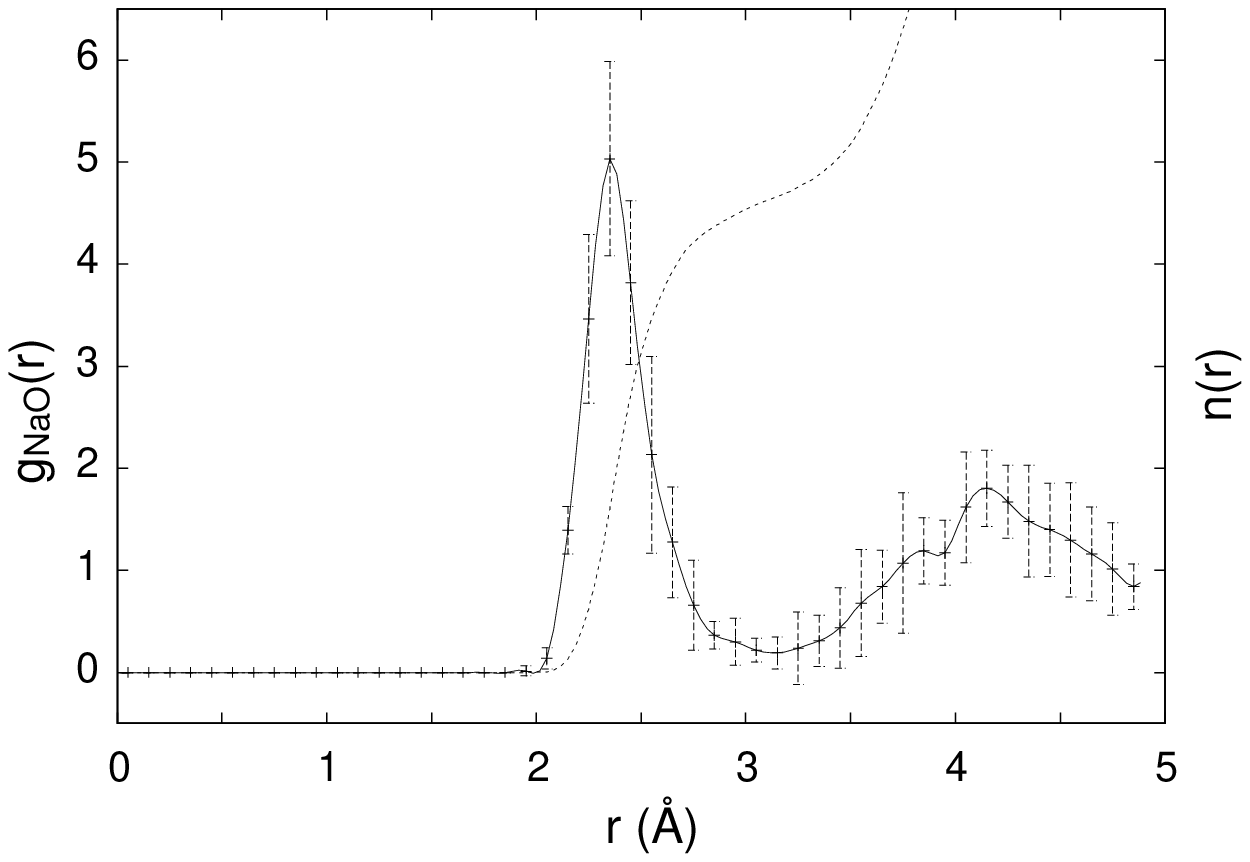}
\end{center}
\caption{Radial distribution function g$_{\mathrm{NaO}}$(r) and number
n(r) of oxygen atoms neighboring the Na$^+$ ion.  Error estimates of
$\pm$ 2$\sigma$ are also plotted for the radial distribution function.
$\sigma$ was estimated by dividing the observed trajectory into four
blocks of approximate duration 1.5~ps; those blocks were assumed to
provide independent observations. The first minimum in the g(r)
function is at r=3.12~\AA\ where g(r) falls to 0.2. Here an average of
4.6 oxygen atoms surround the Na$^+$ ion.}
\label{gor:fig}
\end{figure}

\section{Results}
\begin{figure}[b!]
\begin{center}
\leavevmode
\includegraphics[scale=0.65]{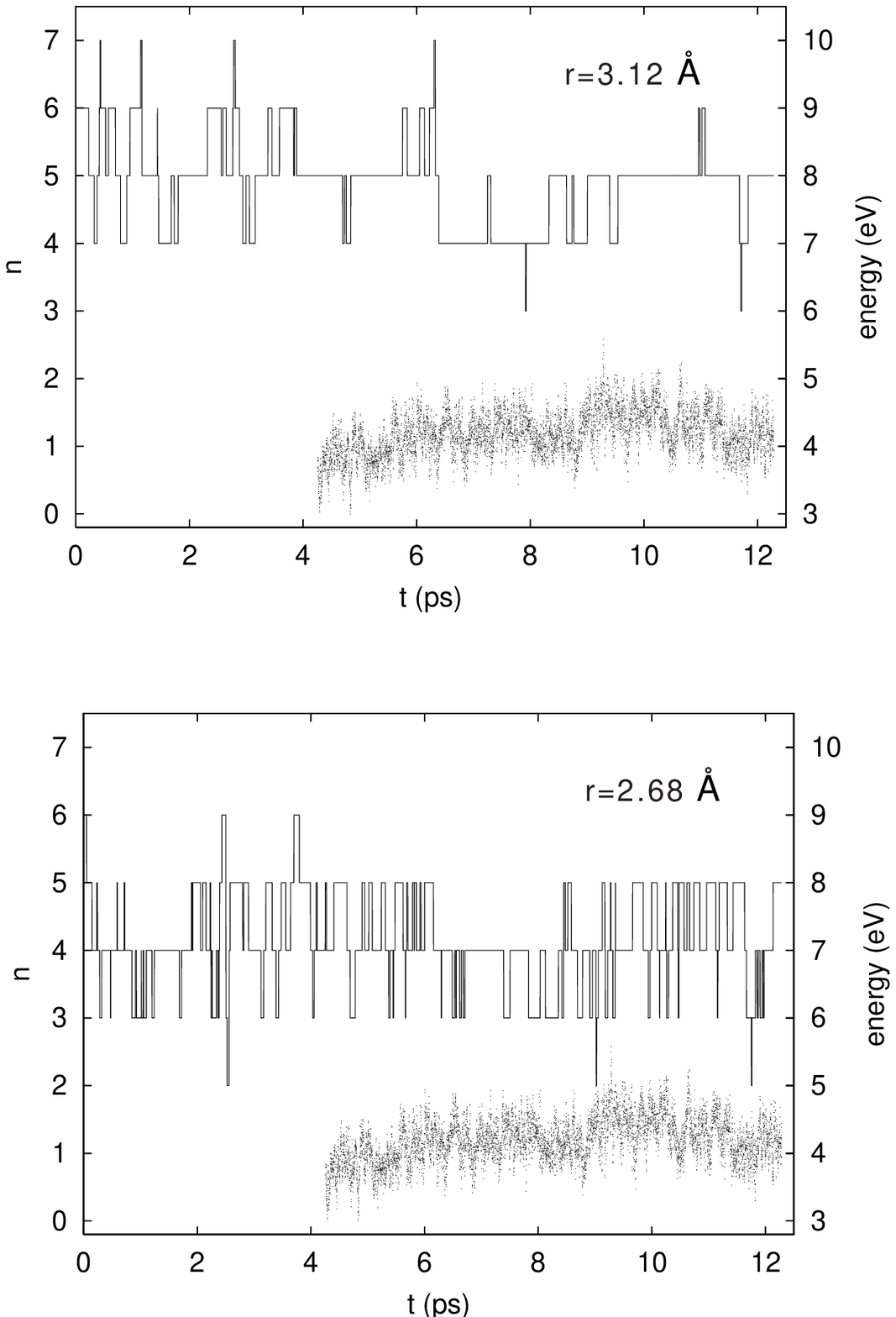}
\end{center}
\caption{The solid line in the upper plot depicts the number of oxygen
atoms within a radius of 3.12 {\AA} from the Na$^+$ at each
configuration in the molecular dynamics simulation.  A radius of 2.68
{\AA} defines the nearest oxygen neighbors in the lower plot.  The
dashed lines show the kinetic energy per atom during the simulation,
plotted after removal of the 300~K thermostat at 4.3~ps. The axis on
the right refers to the kinetic energy values.  In the upper plot, an
average of 4.6 water molecules surround the Na$^+$ ion, while an
average of 4.0 water molecules surround the ion in the lower plot.}
\label{neigh:fig1}
\end{figure}

\begin{figure}[t!]
\begin{center}
\leavevmode
\includegraphics[scale=0.65]{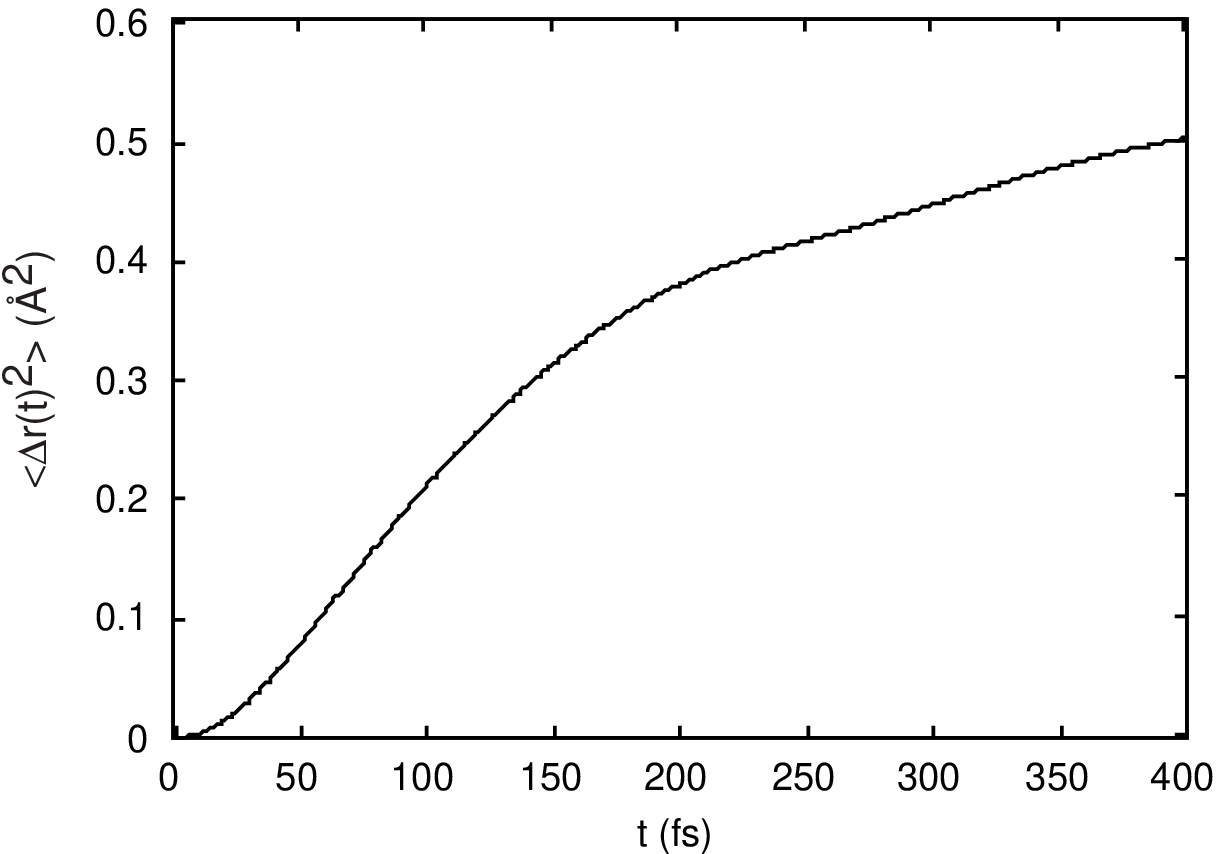}
\end{center}
\caption{Mean-square displacement of the Na$^+$ ion plotted with respect
to the time interval analyzed.  Analysis of the slope from 200-400 ps gives
a diffusion constant of 1.0$\times$10$^{-5}$ cm$^2$/s.} 
\label{diff:fig}
\end{figure}

The ion-oxygen radial distribution function is shown in
Fig.~\ref{gor:fig}. The first maximum occurs at a radius of 2.35~\AA\
from the Na$^+$ ion and the minimum at radius 3.12~\AA\ demarcates the
boundary of the first and innermost hydration shell. An average of
$\langle n\rangle$=4.6  water molecules occupy the inner hydration
shell. Fig.~\ref{neigh:fig1} tracks the instantaneous number of water
oxygen atoms found within the first hydration shell of the Na$^+$,
defined by radius r$\leq$3.12~\AA\ for the upper panel. The fractions
$x_4$ and $x_5$ of four-coordinate and five-coordinate hydration
structures, respectively, constitute $x_4$=40\% and $x_5$=56\% of the
last 6 ps of the simulation. Structures in which the Na$^+$ ion
acquires six innershell water molecules occur with a 4\% frequency,
while structures with three and seven innershell water molecules occur
less than 1\% of the time.  Analysis of the mean-square displacement
of the Na$^+$ ion (Fig.~\ref{diff:fig}) produces a self-diffusion
constant of 1.0$\times$10$^{-5}$ cm$^2$/s, which agrees reasonably
well with an experimental result of 1.33x10$^{-5}$cm$^2$/s\cite{hertz:73}.

These results correspond  coarsely with an `ab initio' molecular
dynamics calculation on this system carried-out
independently\cite{schwegler:00}. The most probable inner shell
occupancy found there was also five, but the probabilities of n=4 and
n=6 were reversed from what we find here.   This difference may be
associated with the lower temperature used in Ref~\cite{schwegler:00}.

One motivation for this study arises from the quasi-chemical theory of
solutions.  According to this formulation, $x_0$ contributes a `chemical'
contribution to $\mu_{Na^+}^{ex}$, the excess chemical potential or
absolute hydration free energy of the ion in liquid
water\cite{pratt:99},
\begin{eqnarray} 
\beta\mu_{\mathrm{Na}^+}{}^{ex}  = \ln x_0 - \ln\left[ \left\langle {e^{-\beta\Delta
U}}\prod\limits_j {(1-b_{\mathrm{Na}^+ j})} \right\rangle_0  \right].
\label{gqca} 
\end{eqnarray}
Here the inner shell is defined by specifying a function
$b_{{\mathrm{Na}^+} j}$ that is equal to one (1) when solvent molecule
j is inside the defined inner shell and zero (0) otherwise;  $\Delta
U$ is the interaction energy of the solvent with the solute Na$^+$
that is treated as a test particle, $\beta^{-1}$=k$_B$T, and the
subscript zero associated with $ \left\langle \ldots \right\rangle_0$
indicates a test particle average \cite{pratt:99}. The second term on
the right-hand side of Eq.~(\ref{gqca}) is the excess chemical
potential of the solute lacking inner shell solvent molecules whereas
the first term is the free energy of allowing solvent molecules to
occupy the inner shell. The  validity of Eq.~(\ref{gqca}) has been
established elsewhere\cite{pratt:99}.  The second term on the right of
Eq.~\ref{gqca} is the outer sphere contribution to the excess chemical
potential in contrast to the first or chemical term.

The utility of this quasi-chemical formulation is the
suggestion\cite{hummer:cp:2000} of more detailed study of the $x_n$,
the fractions of n-coordinate hydration structures found in solution,
on the basis of the equilibria forming inner shell complexes of
different aggregation number:
\begin{eqnarray} \mathrm{Na(H_2O)_{m=0}{}^+}\ +\ \mathrm{nH_2O}
\rightleftharpoons \mathrm{Na(H_2O)_n{}^+}~.
\label{reaction} \end{eqnarray}
Utilizing the chemical equilibrium ratios
\begin{eqnarray}
K_n={\rho_{\mathrm{Na(H_2O)_n}{}^+} \over
\rho_{\mathrm{H_2O}}{}^n\rho_{\mathrm{Na(H_2O)_{m=0}{}^+}}
 }~,
\label{Kn-ob}
\end{eqnarray}
the normalized $x_n$ can be expressed as  
\begin{eqnarray}
x_n={{K_{n}\rho _{\mathrm{H_2O}}{}^n} \over {\sum\limits_{m\ge 0}
{K_{m}\rho _{\mathrm{H_2O}}{}^m}}}~.
\label{cluster-var}
\end{eqnarray}
The $\rho_\sigma$ are the number densities and, in particular,
$\rho_{\mathrm{H_2O}}$ is the molecule number density of liquid water.
If the medium external to the clusters is neglected, the equilibrium
ratios, denoted as $K_n{}^{(0)}$, can be obtained from electronic
structure calculations on the complexes, assuming for the thermal
motion of the atoms the harmonic approximation evaluated at the
calculated minimum energy configuration.  Finally utilization of a
dielectric continuum approximation for the outer sphere contributions
to the chemical potential gives a natural, though approximate,
quasi-chemical
model\cite{rempe:99,pratt:98,feature,martin:97,pratt:99,G98}.

For the present problem, the quasi-chemical approximation was implemented
following precisely the procedures of the earlier study of
Li$^+$(aq)\cite{rempe:99}, except that the sodium ion cavity radius 
for the dielectric model calculation was assigned as
R$_{Na^+}$=3.1~\AA, the distance of the first minimum of the radial
distribution function of Fig.~\ref{gor:fig}.    The temperature and
density used were 344~K and 1.0~g/cm$^3$ and the value of the bulk
dielectric constant was 65.3\cite{uematsu:80}.

\begin{figure}[b!]
\vspace{0.2in}
\begin{center}
\leavevmode
\includegraphics[scale=0.45]{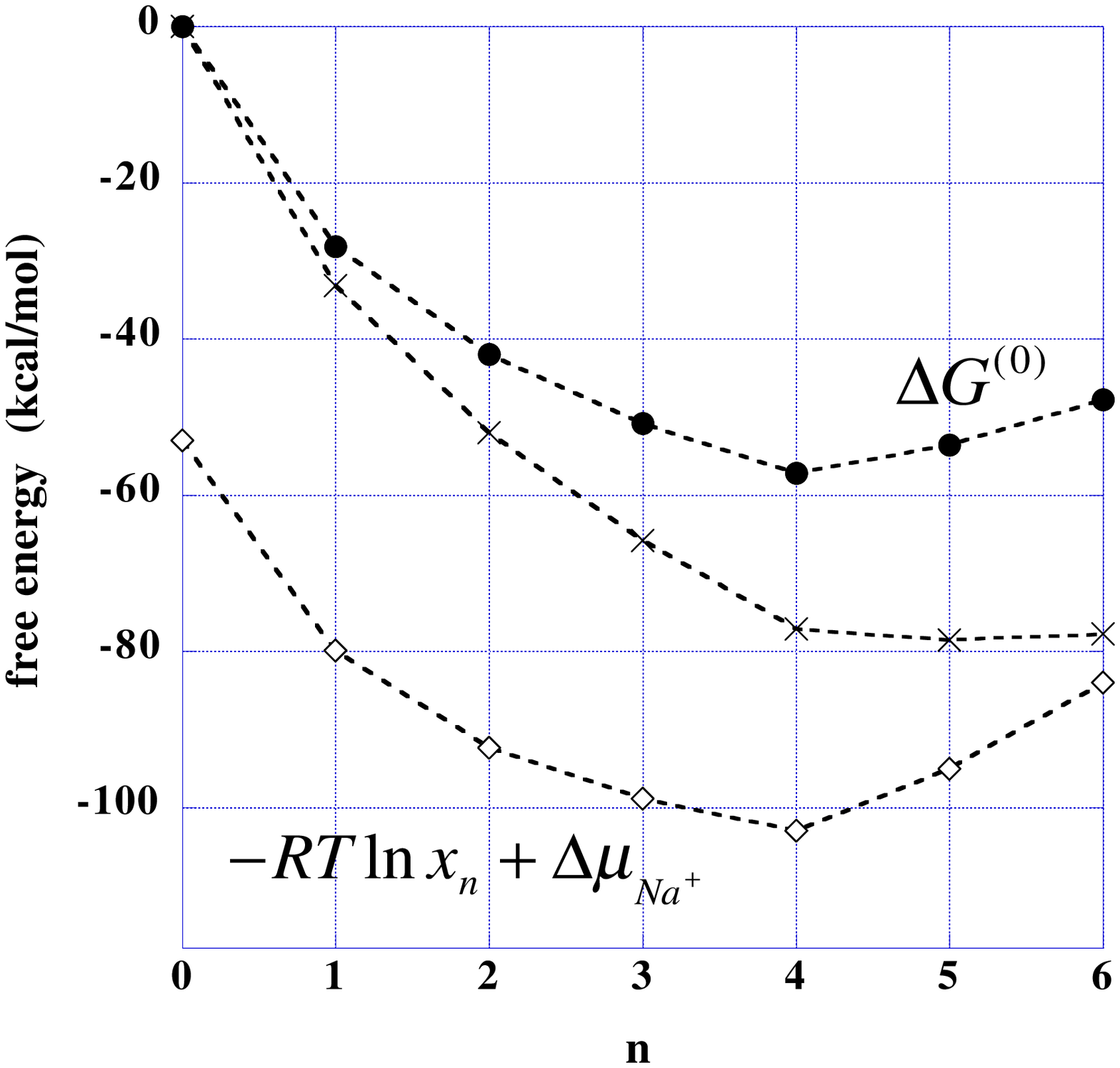}  
\end{center}
\caption{Free energies for Na$^+$ ion hydration in liquid water as a
function of the number of inner shell water neighbors at T=344~K and
$\rho_{\mathrm{H_2O}}$=1~g/cm$^3$.   The lowest results (open
diamonds) show quasi-chemical approximate values for the liquid,
labelled according to the quasi-chemical interpretation. This graph
indicates that the n=4 inner sphere structure is most probable under
these conditions.  The radius used for the Na$^+$ ion here is 3.1~\AA,
though a substantial reduction of this value produced only a minor
change in the inferred absolute hydration free energy; otherwise the
procedure is the same as in previous
reports\protect\cite{rempe:99,pratt:99}. The absolute hydration free
energy predicted here is -103~kcal/mol.  The results marked
$\Delta$G$^{(0)}$ (filled circles) are the free energies predicted for
the reaction Na$^+$ + n~H$_2$O in an ideal gas at p = 1~atm
$\equiv{\bar p}$ and T=344~K.  The minimum value is at n=4. The middle
graph (crosses) add to the ideal gas results the `replacement'
contribution reflecting the formal density of the
water molecules $\mathrm{-n RT}\, \ln \left[ \mathrm{RT}
\rho_{\mathrm{H_2O}} / {\bar p}  \right] = \mathrm{-n*5.03}$
kcal/mol with T=344~K, and $\rho_{\mathrm{H_2O}}$ =1~g/cm$^3$.}
\label{qc:fig}
\end{figure}

Results of the calculations are summarized in Fig.~\ref{qc:fig}. The
electronic structure results are consonant with those found previously
for the Li$^+$ ion.  The n=4 inner sphere gas-phase complex has the
lowest free energy.  Although outer sphere placements are obtained for
additional water molecules in the minimum energy structures of larger
clusters, attention is, nevertheless,  here restricted to inner sphere
structures. The mean occupation number predicted by this
quasi-chemical model is $\langle n \rangle$ = 4.0; the computed
absolute hydration free energy of the Na$^+$ ion under these
conditions is -103~kcal/mol, not including any repulsive force
(packing) contributions. An experimental value  for Na$^+$ ion in
liquid water at room temperature is -87~kcal/mol\cite{marcus:94}.

\begin{figure}[t!]
\begin{center}
\leavevmode
\includegraphics[scale=0.7]{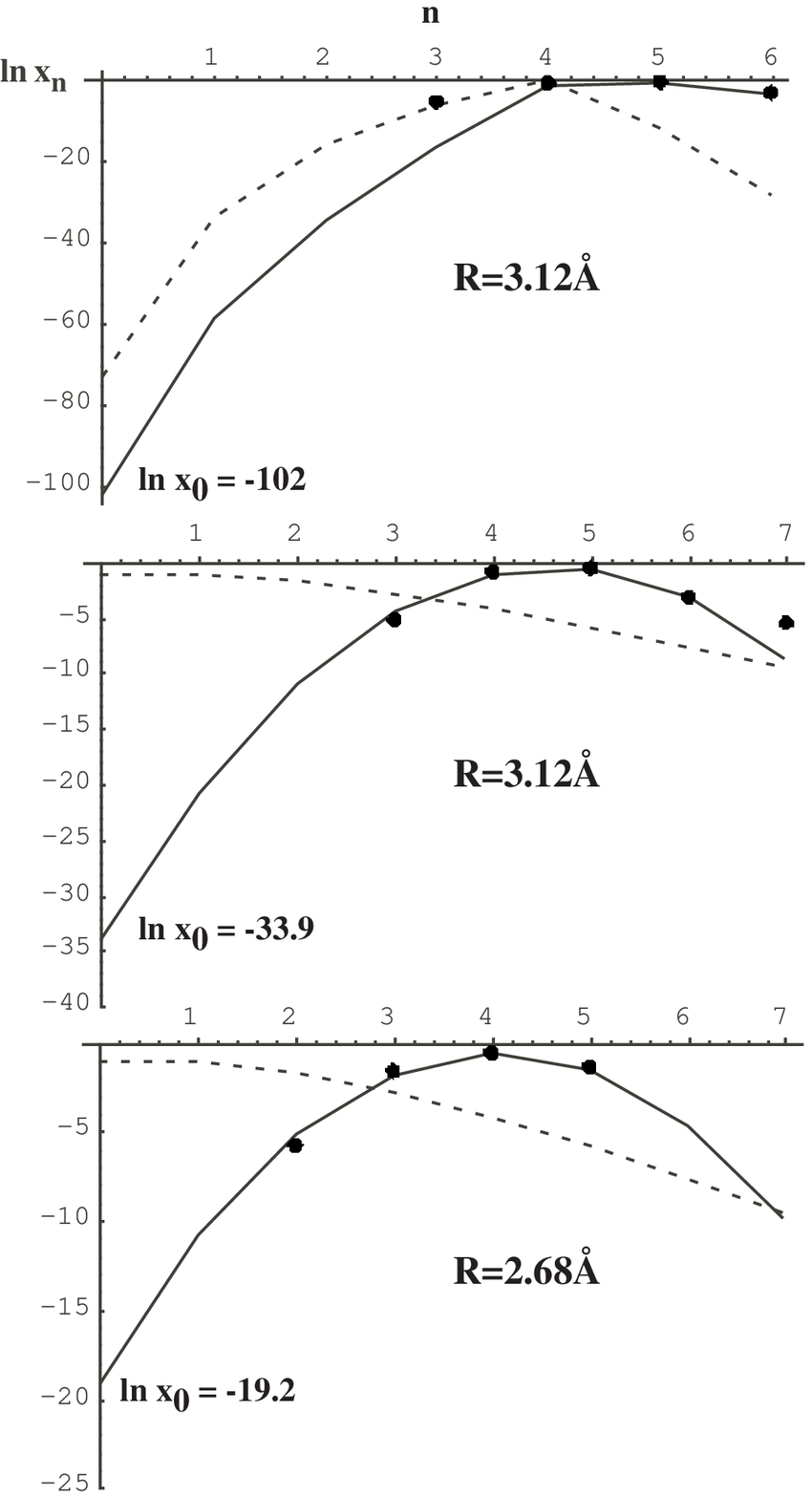}
\end{center}
\caption{Results for the inference of $x_0$ from `ab initio' molecular
dynamics information.  The solid points represent the information
extracted from the molecular dynamics simulation, the dotted lines are
the default models, and the solid lines show the fit achieved by the
information theory approach. In the top panel, the hepta-occupancy was
excluded, the probabilities were renormalized on this condition, and
the quasi-chemical default model was used together with the moments
$\langle n \rangle$=4.633 and $\langle n(n-1)/2 \rangle\rangle$=8.577.
In the middle panel, all $x_n$ observed in the simulation were
included, $\langle n \rangle$=4.642 and $\langle n(n-1)/2
\rangle\rangle$=8.624, and the ideal gas (or Gibbs default) model was
used along with the same moments as above. The bottom panel shows the
results using the inner sphere radius R=2.68~\AA\ and the moments
$\langle n \rangle$=4.046, and $\langle n(n-1)/2 \rangle\rangle$=6.393
with the Gibbs default model. } 
\label{inf:fig}
\end{figure}

Because of the significance of $x_0$ [Eq.~\ref{gqca}], we fitted
several model distributions \{$x_n$\} based on different ideas to the
`ab initio' molecular dynamics results.  The varying success of those
models in inferring $x_0$ was enlightening.  An instructive selection of those models is
shown in Fig.~\ref{inf:fig} and we describe those results here.

First, we note that though the preceeding quasi-chemical approximation
does not agree closely with the 'ab initio' molecular dynamics simulation, the populations
obtained from the quasi-chemical approximation, $\hat x_n$, can serve
as a default model for a maximum entropy inference of
$x_n$\cite{hummer:cp:2000}. In this approach we model
\begin{equation}
\ln x_j=\ln \hat x_j -\lambda _0-j\lambda _1-j(j-1)\lambda _2/2 -
\ldots~,
\label{maxent}
\end{equation}
with Lagrange multipliers  $\lambda_k$  adjusted to conform to
available moment information
\begin{equation}
\left\langle {n \choose j} \right\rangle =\sum\limits_k {x_k}{k
\choose j}
\label{moments}
\end{equation}
for j = 0, 1, 2,  \ldots.  In view of the limited data available, use
of more than two moments produced operationally ill-posed fitting
problems.

One difficulty with this specific approach is that the `ab initio'
molecular dynamics produced $x_7\!\!>$0 in contrast to the electronic
structure methods that found no minimum energy 
hepta-coordinated inner-sphere clusters.  Since
the observed $x_7$  is likely to be relatively less accurate and is
furthest away from the desired n=0 element, we excluded n=7
configurations of the `ab initio' molecular dynamics, renormalized
the probabilities $x_n$, and recalculated the moments.
As the upper panel in Fig.~\ref{inf:fig} shows, this simple maximum
entropy model is qualitatively satisfactory although not
quantitatively convincing. The fitted model significantly disagrees
with the observed $x_3$.  The chemical contribution suggested by
Fig.~\ref{inf:fig} is approximately -70~kcal/mol. Using the Born
formula, $-q^2(1 - 1/\epsilon)/2R$ with R=3.12~\AA, to estimate the
outer sphere contributions represented by the last term in
Eq.~\ref{gqca}, then the net absolute hydration free energy falls in
the neighborhood of -115~kcal/mol.  Since experimental values for the
absolute hydration free energy at room temperature center around
-90~kcal/mol, this comparison shows that the present free energy
results are not to be interpreted quantitatively, but rather as
indicative of the present state of the theory.

A second approach focused on testing a default model that supplies a
nonzero ${\hat x}_7$; we used the Gibbs default model $\hat{x}_n
\propto 1/n!$ that would give the correct answer for an ideal gas
`solvent.'  This model has the additional and heuristic advantage of being
significantly broader.  Our experience has been that these maximum
entropy fitting procedures work better when the default model is
broader than the distribution sought. The results, illustrated in the
middle panel of Fig.~\ref{inf:fig}, show an improved fit.  Here the
chemical contribution to the free energy is -23~kcal/mol, yielding a
net absolute hydration free energy of -68~kcal/mol when the same Born
formula is used to estimate the outer sphere contributions.

A third fitting possibility was based on a suggestion from a previous
`ab initio' molecular dynamics calculation on K$^+$(aq): that the
innermost {\em four} water molecules have a special
status\cite{ramaniah:98}.  In fact, the quasi-chemical approximation above
and the fitting of the upper panel of Fig.~\ref{inf:fig} suggests also
that the $x_n$ results for n$\le$4 and for n$\ge$5 display
different behaviors. The radial distribution
function of Fig.~\ref{gor:fig}, somewhat better resolved than
heretofore, is relevant to this issue and, in contrast, doesn't
directly support a hypothesis of two populations of water molecules in
the inner shell. Nevertheless, that g(r) does not rule out the possibility
that the structures might become more flexible as the inner shell
nears maximum capacity with lower incremental binding energies.

To clarify these possibilities, we reduced the radius defining the
inner sphere to R=2.68~\AA, for which $<$n$>$ is close to 4 (see
bottom panel of Fig.~\ref{gor:fig}) and reanalysed the `ab initio'
molecular dynamics trajectory to extract the appropriate alternative
moment information. Again using the Gibbs default model, we obtained
the results shown in the lowest panel of Fig.~\ref{inf:fig}.  The
inferred chemical contribution is  $RT\ln x_0 \approx$~-13~kcal/mol. 
Using again the Born approximation for the outer sphere
contribution, this time with R=2.68~\AA, we obtain an absolute
hydration free energy estimate of -65~kcal/mol.

\begin{figure}[tb!]
\begin{center}
\leavevmode
\includegraphics[scale=0.7]{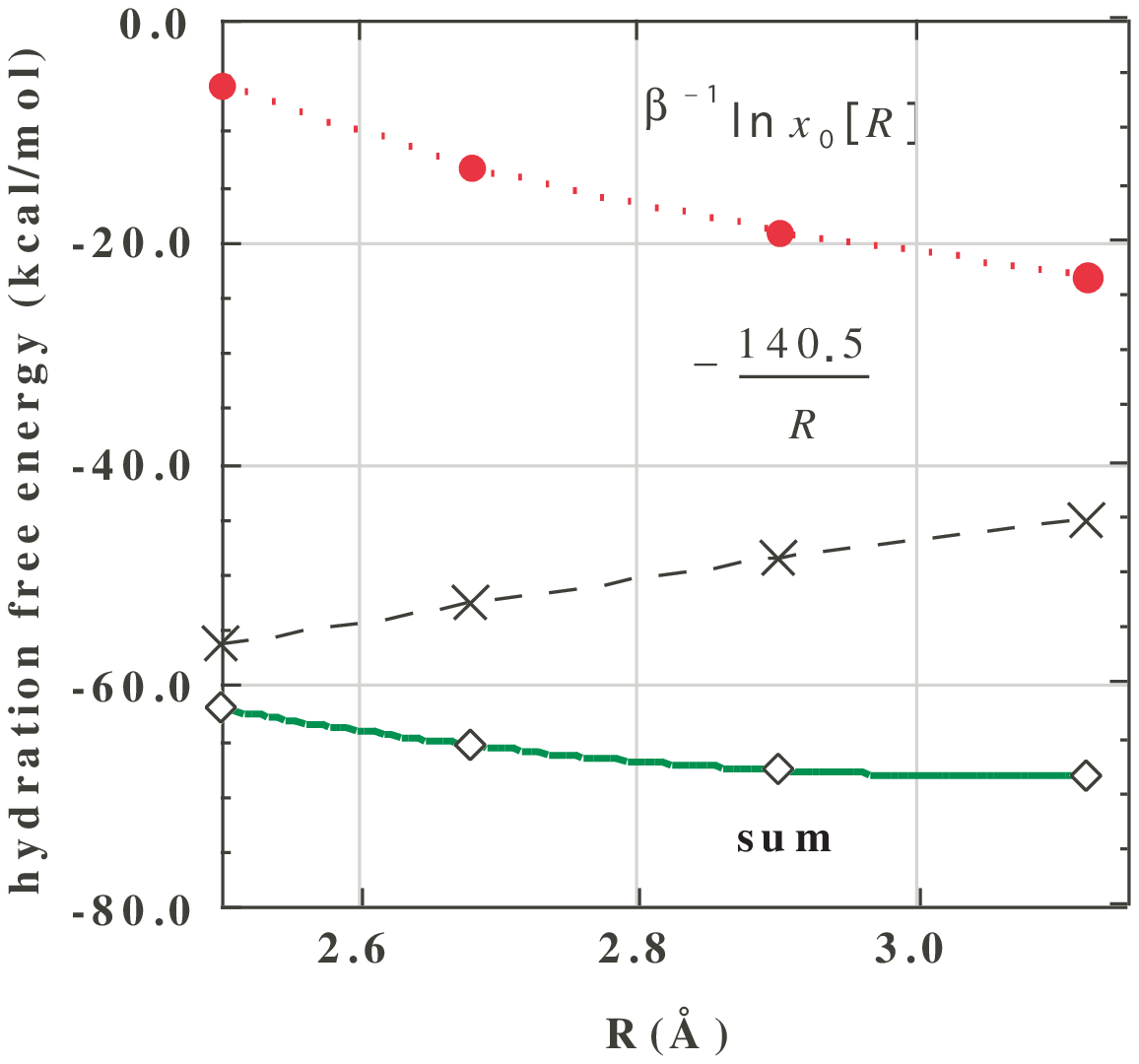}
\end{center}
\caption{Variation of hydration free energy contributions with changes
in radius R defining the inner sphere.   The upper curve is
the chemical contribution obtained with an information theory fit
using a Gibbs default. The middle curve is the outer sphere
contribution approximated by the Born formula for a spherical ion with
unit charge at its center. The bottom curve is the sum of the other
two.  The slope of the bottom curve is zero at R$\approx$3.06\AA.  As
discussed in the text, this value identifies a R-region for which the
approximations used are pragmatically balanced.  The curve takes the
value -68~kcal/mol in that region.}
\label{twomcut:fig}
\end{figure}

The insensitivity of these latter results to choice of inner sphere
radius deserves emphasis and further discussion.  From a formal point
of view, the inner sphere radius R serves only a bookkeeping role;
the left side of Eq.~\ref{gqca} should be strictly unaffected by
changes in R.  Nevertheless, the terms on the right side of that
equation are individually affected by changes in R.  Thus, we might
take the insensitivity of the sum of those individual terms as an
indication that the inevitable approximations are reasonably balanced.
Values of R for which the sum  Eq.~\ref{gqca} is insensitive are
pragmatic values given the approximations made. 
Fig.~\ref{twomcut:fig} illustrates these points and establishes the
pragmatic value R$\approx$3.06\AA\ for the current application.  The
similarity of this value with the radius of the inner shell suggested
by Fig.~\ref{gor:fig} (3.12\AA) is encouraging.  The value -68~kcal/mol
is then suggested for the hydration free energy in the absence of any
account of packing or van der Waals interactions.

\section{Conclusions} The `ab initio' molecular dynamics simulation
predicts the most probable occupancy of the inner shell of Na$^+$(aq)
to be 5 and the mean occupancy to be 4.6 water molecules at infinite
dilution, T=344~K, and a nominal water density of 1~g/cm$^3$.  The
simulation produces both a satisfactory Na-O radial distribution
function and self-diffusion coefficient for Na$^+$, but these
satisfactory results required more care with thermalization and
averaging time than is most common with these demanding calculations.  Recently,
this point has been separately emphasized in the context of `ab initio'
simulation of water\cite{sorenson:00}.

The complementary calculation framed in terms of quasi-chemical theory
based on electronic structure results for ion-water clusters, the
harmonic approximation for cluster motion,  and a dielectric continuum
model for outer sphere contributions underestimates the inner shell
water molecule occupancies for Na$^+$ in liquid water. Maximum entropy
fitting of the inner shell occupancy distribution shows that the
ion-water cluster results yield a distribution significantly narrower
than that obtained from the simulations. For this reason, naive
inference of the absolute hydration free energy of Na$^+$(aq) based on
the cluster electronic structure results and utilizing information
gleaned from the `ab initio' molecular dynamics was unsuccessful.  The
electronic structure calculations found minimum energy
Na(H$_2$O)$_7{}^+$ clusters only with obvious outer sphere placements
of some water molecules though hepta-coordinate inner sphere clusters were
observed in the `ab initio' molecular dynamics with the most natural
cluster definition. These results suggest that the anharmonicities and
large amplitude motion are serious concerns, particularly for the
larger clusters, and that the approximate theory utilized for outer
sphere contributions must treat cluster conformations differently
from minimum energy structures of the isolated clusters.

Abandonment of the cluster electronic structure results in favor of a
broader default model improved the modeling of the $x_n$ distribution on
the basis of the information extracted from the simulation. A sequence
of more aggressive fits eventually suggested the value -68~kcal/mol for the
hydration free energy at this somewhat elevated temperature on the
basis of the quasi-chemical perspective of inner sphere occupancies but in the
absence of any account of packing or van der Waals interactions.

We acknowledge helpful discussions of many related issues with Gerhard
Hummer and Joel Kress. This work was supported by the US Department of
Energy under contract W-7405-ENG-36 and the LDRD program at Los
Alamos.


%
%

%


\bibliography{your bib file}

\end{document}